\documentclass[useAMS,usenatbib]{mn2e}
\usepackage{natbib}
\usepackage{graphicx}
\usepackage{aalongtable}
\usepackage{txfonts}

\begin{document}

\title[A statistical study of gaseous environment of {\it Spitzer} interstellar bubbles]
      {A statistical study of gaseous environment of {\it Spitzer} interstellar bubbles}

\author[L.~G. Hou and X. Y. Gao]
{L.~G. Hou\thanks{E-mail: lghou@nao.cas.cn}
and X. Y. Gao
\\
National Astronomical Observatories, Chinese Academy of Sciences, 20A
Datun Road, Chaoyang District, Beijing 100012, China\\
}

\date{Accepted 2013 ... Received 2013 ...}

\pagerange{\pageref{firstpage}--\pageref{lastpage}} \pubyear{2013}

\maketitle

\label{firstpage}

\begin{abstract}

The expansion of interstellar bubbles is suggested to be an important
mechanism of triggering material accumulation and star formation.
In this work, we investigate the gaseous environment of a large sample
of interstellar bubbles identified by the {\it Spitzer} space
telescope, aiming to explore the possible evidence of triggered gas
accumulation and star formation in a statistical sense.
By cross-matching 6\,124 {\it Spitzer} interstellar bubbles from the
Milky Way Project (MWP) and more than 2\,500 Galactic H\,II regions
collected by us, we obtain the velocity information for 818 MWP
bubbles. To study the gaseous environment of the interstellar bubbles
and get rid of the projection effect as much as possible, we constrain
the velocity difference between the bubbles and the $^{13}$CO(1$-$0)
emission extracted from the Galactic Ring Survey (GRS). Three methods:
the mean azimuthally averaged radial profile of $^{13}$CO emission,
the surface number density of molecular clumps, and the angular
cross-correlation function of MWP bubbles and the GRS molecular clumps
are adopted. Significant over density of molecular gas is found to be
close to the bubble rims. 60\% of the studied bubbles were found to
have associated molecular clumps. By comparing the clump-associated
and the clump-unassociated MWP interstellar bubbles, we reveal that
the bubbles in associations tend to be larger and thicker in physical
sizes. From the different properties shown by the bubble-associated
and bubble-unassociated clumps, we speculate that some of the
bubble-associated clumps result from the expansion of bubbles. The
fraction of the molecular clumps associated with the MWP bubbles is
estimated to be about 20\% after considering the projection
effect. For the bubble-clump complexes, we found that the bubbles in
the complexes with associated massive young stellar object(s)
(MYSO(s)) have larger physical sizes, hence the complexes tend to be
older. We propose that an evolutionary sequence might exist between
the relative younger MYSO-unassociated bubble-clump complexes and the
MYSO-associated complexes.

\end{abstract}


%
%________________________________________________________________

\section{Introduction}
\label{sect:intro}

Massive stars have significant dynamical influences on the ambient
molecular clouds (MCs) by several feedback mechanisms, i.e.  H\,II
regions, outflows, stellar winds, jets, and supernovae explosions,
among which the expansion of H\,II regions is suggested to be one of
the most important factors~\citep{matz02,deb12}. The expanding H\,II
regions reshape the surrounding interstellar medium and may trigger
the star formation (SF)~\citep[e.g.][]{wpc+08,dsa+10,tumm12}. These
effects are predicted in theories and searched in observations.

Theoretically, several mechanisms of triggering star formation by
H\,II regions have been proposed. Two major mechanisms are the
``collect \& collapse'' process~\citep{el77} and the ``radiation
driven implosion'' (or ``radiation driven compression'') of
pre-existing dense molecular clumps~\citep{ll94,dsa+10}. Other
mechanisms such as ionizing radiation acting on MCs with turbulent
gas~\citep{ekt95} and dynamical instabilities of ionizing
front~\citep{gf96} have also been discussed. These mechanisms do not
conflict with each other, and can work simultaneously in a single
H\,II region~\citep{dzc05}.

Observationally, evidence for triggered SF by the expansion of H\,II
regions was searched by studying the distribution of neutral material,
young stellar objects (YSOs), and also the ages of molecular clumps or
YSOs related to their associated H\,II regions. Recently, such topic
has extensively benefited from the release of a large number of
interstellar bubbles\footnote{The word ``bubbles'' has also been used
  by \citet{mkf+10}, who identified 416 disk- and ring-like objects
  from the {\it Spitzer/}MIPSGAL 24 $\mu$m survey. Their ``bubbles''
  are small in the apparent size and most of them are related with the
  circumstellar envelopes of evolved stars.} discovered in the {\it
  Spitzer}/GLIMPSE survey~{(Churchwell et al. 2006, 2007, hereafter
  CH06\&07; Simpson et al. 2012)}. A {\it Spitzer} interstellar bubble
is clearly characterized by its ring- or arc-shaped structure shown in
the GLIMPSE 8 $\mu$m image. The 8~$\mu$m band of {\it Spitzer}$-$IRAC
is dominated by the emission from the polycyclic aromatic hydrocarbon
(PAH) molecules or clusters of molecules~\citep{wpc+08}. These species
are excited by absorbing the far-UV photons leaking from the central
ionized gas. Hence, their emissions trace the ionizing fronts of H\,II
regions. The 8 $\mu$m ring- or arc-shaped structure encircles the 24
$\mu$m emission features tracing the hot dusts embedded in the ionized
gas~\citep{wpc+08}, and also encircles the radio free-free emission
features indicating the central ionized gas~\citep[see e.g., the Fig.3
of][]{dsa+10}. Hence, the observational properties of the {\it
  Spitzer} interstellar bubbles are in good agreement with that
  of expanding H\,II regions~\citep{dsa+10}. By visual identification
of YSO(s) and neutral material near the interstellar bubbles, the
possible evidences of triggered SF and/or material accumulation were
reported towards several individual sources, e.g. bubbles N\,10 and
N\,49~\citep{wpc+08,zar+10,dib+12}, H\,II region
RCW\,120~\citep{dzs+09}, and star forming region
M\,17~\citep{pcb+09}. Similar studies were made towards some other
targets, i.e. W\,51A~\citep{kbkl09}, G45.45+0.06~\citep{pco09},
RCW\,79, RCW\,82, RCW\,120~\citep{mpd+10}, RCW\,34~\citep{bpw+10},
bubble N\,65~\citep{ppg10}, S\,51~\citep{zw12a},
N\,22~\citep{jze+12,sher12}, Sh2-294~\citep{spo+12},
N\,68~\citep{zw13}, N\,14~\citep{sher12,do12}, N\,74~\citep{sher12},
N\,131~\citep{zwx13}, G041.10$-$0.15, G041.91$-$0.12, N\,80, N\,91,
N\,92~\citep{dib+12}, N\,4~\citep{ljlw13} and
Sh2-284~\citep{phn+09}. However, as pointed out by~\citet{tumm12},
these methods are almost impossible to diagnose the origin of the
identified YSOs. Because it is always hard to distinguish the real
physical association from the projection effect. Furthermore, with a
single time observational data towards one system, it is also very
difficult to exclude the possibility that some or even all of the
identified YSO(s), protostar(s) and molecular clump(s) were formed
spontaneously without any influences of triggering. As shown by the
simulation of \citet{deb12b}, the association of stars with the shell
or pillar-like structures in the gas is a good but not a predominant
indication of triggering.

Besides the investigations on individual objects, statistical studies
of the correlations between large samples of YSOs or gases with
interstellar bubbles provide a different prospect and shed light on
inferring the presence of triggering in a statistical
sense. \citet{tumm12} first made a statistical study of the massive
star formation around 322 interstellar bubbles from CH06\&07. Using
the massive YSOs (MYSOs) identified from the Red MSX Source (RMS)
survey~\citep{uhl+08}, they found a significant over density of MYSOs
near the bubble rims. \citet{ksb+12} made a similar analysis by using
a more complete bubble catalogue from the Milky Way Project
(MWP)~\citep{spk+12}. They found a strong positional correlation of
the RMS MYSOs/H\,II regions with bubbles at angular distances less
than two effective bubble radii. Both of the two studies imply that a
significant fraction of YSOs {\it projectively} associated with the
bubble rims are {\it likely} triggered by the expansion of the
bubbles/H\,II regions. To confirm this hypothesis, a statistical study
on the gaseous environment of the interstellar bubbles is definitely
needed.

Other than the distribution of YSOs around the bubbles, the
observational evidence for triggered SF and material accumulation
should be also present for neutral gases. The obstacle in such
studies, as mentioned above is how to solve the projection effect in
distinguishing the true association between the interstellar bubbles
and the gaseous components. Previous efforts concentrated on visual
inspection of channel maps of the observed molecular lines toward {\it
  individual} bubbles, through which morphological similarities
between the bubbles and the ambient gas are expected. A better
solution that we propose in this work might be taking the velocity
information into account, since similar line-of-sight velocities of a
bubble and its physically associated gas components are expected. Such
attempt in a statistical sense now becomes available due to the
presence of the 5\,106 {\it Spitzer} interstellar bubbles identified
by the MWP~\citep[][]{spk+12}, the largest catalogue of about 2\,500
Galactic H\,II regions with velocity measurements compiled by us (Hou
\& Han 2013, in prep., to be submitted) and the publicity of a number
of surveys of molecular gas~\citep[e.g.][]{dht01,grs06}. We aim to
statistically investigate the gaseous environment of the interstellar
bubbles and explore the possible evidence of triggered material
accumulation and star formation related to the bubbles. This work is
organized as follows: In Sect. 2, we introduce the data sets of the
interstellar bubbles and the molecular gas. In Sect. 3, the gaseous
environment of bubbles are explored by three different
methods. Discussions and conclusions are given in Sect. 4 and Sect. 5,
respectively.

%__________________________________________________________________

\section{Data}

To statistically study the gaseous environment of the interstellar
bubbles and get rid of the projection effect as much as possible, a
large sample of bubbles with line-of-sight velocity information and a
systematic survey of molecular gas with appropriate resolution and
sampling are needed.

\subsection{{\it Spizter} interstellar bubbles with velocity information}
\label{velocity}

The interstellar bubble catalogue adopted in this work is from the
recently released Milky Way Project~\citep[][]{spk+12}, containing
5\,106 bubbles identified by visual inspection of the {\it
  Spitzer}/GLIMPSE~\citep{bcb+03} and MIPSGAL~\citep{cnm+09} survey
images. This catalogue has a sky coverage of $|l|<65^\circ$ and
$|b|<1^\circ$. It increases the number of known interstellar bubbles
\citep{cpa+06,cwp+07} by nearly an order of magnitude. Because the
observational properties of the {\it Spitzer} interstellar bubbles are
in good agreement with that of expanding H\,II
regions~\citep{dsa+10}, to obtain the velocity information of the
bubbles, we rely on the cross match between the bubbles and the H\,II
regions whose line measurements (e.g. radio recombination lines,
molecular lines) have been made.

\begin{figure}
\centering\includegraphics[width=0.5\textwidth]{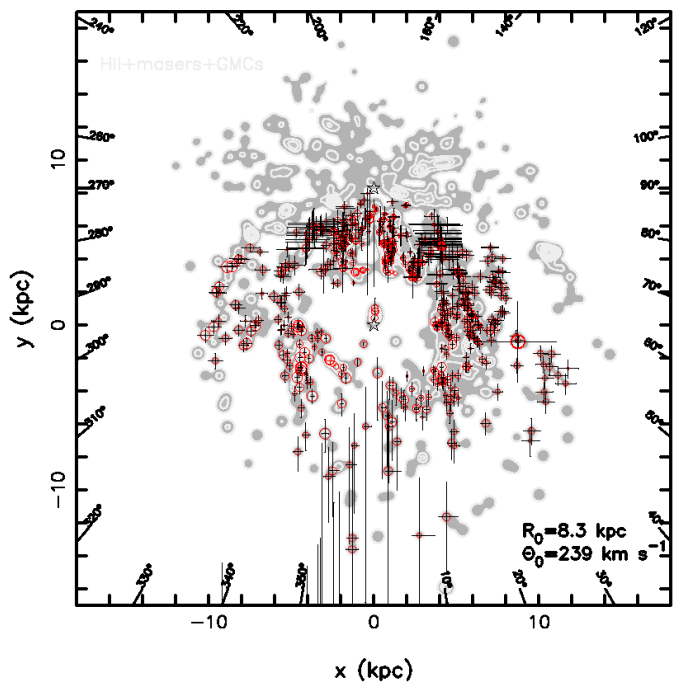}
\caption{Projected distribution of bubbles (red circles) on the
  Galactic plane. Background is an ``intensity'' distribution of
  spiral tracers \citep[see][for a detail]{hhs09,hh13b}. Two stars
  represent the location of the Sun (x~=~0.0~kpc, y~=~8.3~kpc) and the
  Galactic center (x~=~0.0~kpc, y~=~0.0~kpc). A flat rotation curve
  with R$_0$~=~8.3~kpc and $\theta_0$~=~239~km~s$^{-1}$ is used to
  calculate the kinematic distances if no photometric or trigonometric
  distances are available. The error bars are estimated by considering
  a velocity uncertainty of $\pm$ 7 km s$^{-1}$ if no errors are given
  by the trigonometric or photometric observations.}
\label{dist_g}
\end{figure}
\begin{figure*}
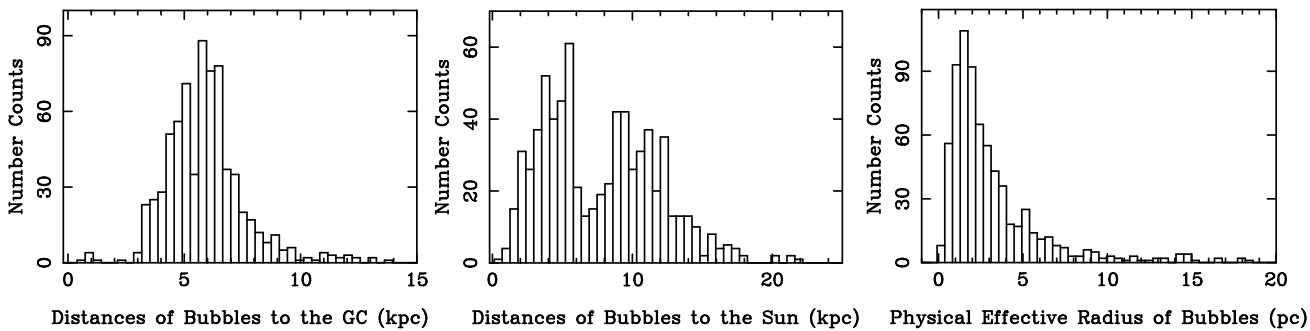

\centering\includegraphics[width=0.316\textwidth]{dist_GC.ps}
\centering\includegraphics[width=0.32\textwidth]{dist_sun.ps}
\centering\includegraphics[width=0.332\textwidth]{bubble_size.ps}
\caption{Distributions of bubble distances to the Galacitic center
  ({\it left panel}) and to the sun ({\it middle panel}). {\it Right
    panel}: Distribution of the physical effective radius of
  bubbles. A flat rotation curve with R$_0$~=~8.3~kpc and
  $\theta_0$~=~239~km~s$^{-1}$ is used to calculate the kinematic
  distances if no photometric or trigonometric distance is available.}
\label{dist}
\end{figure*}

Aiming at illustrating the spiral structure of the Milky Way,
\citet{hhs09} compiled a large sample of Galactic H\,II regions. By
collecting the new measurements released since then, a catalogue
including about 2\,500 Galactic H\,II regions and confident candidates
with more than 4\,300 line measurements is compiled (Hou \& Han 2013,
in prep.). It is by far the largest sample of Galactic H\,II regions
with velocity and distance information. 1\,677 of them fall within the
sky coverage of the MWP.

A cross match is first made between the MWP interstellar bubbles and
the Galactic H\,II regions. If the angular separation between a bubble
and an H\,II region is less than one effective radius\footnote{Here,
  the effective radius is defined as
  $R_{eff}=(R_{in}r_{in})^{0.5}/2+(R_{out}r_{out})^{0.5}/2$, where
  $R_{in}$ and $R_{out}$ are the inner and outer semi-major axes of a
  bubble, and $r_{in}$ and r$_{out}$ are the inner and outer
  semi-minor axes of a bubble, respectively~\citep{spk+12}.} of the
bubble, they are taken as coincident. 818 out of the 5\,106 MWP
bubbles were finally found to have matched H\,II regions (see details,
e.g. coordinates, velocities in Table~\ref{tab_1}). Among them, 721
H\,II regions have published distances. This number is nearly four
times larger than previously known~\citep[189 in][]{ksb+12}. We
directly assign the distance to a bubble if its corresponding H\,II
region has a photometric or trigonometric distance. Otherwise, their
published kinematic distances are re-calculated based on the new
parameters: first, we modify the V$_{LSR}$ according to the new solar
motions~\citep[U$_0$~=~10.27~km~s$^{-1}$, V$_0$~=~15.32~km~s$^{-1}$
  and W$_0$~=~7.74~km~s$^{-1}$,][]{sbd10}, the new values of
R$_0$~=~8.3~kpc and $\theta_0$~=~239~km~s$^{-1}$ are then used for a
flat Galactic rotation curve \citep[][]{brm+11}.

The distribution of the 721 MWP bubbles whose distances are obtained
from their associated H\,II regions is presented in
Fig.~\ref{dist_g}. They spread throughout the Galactic plane. The
distances to the Galactic center of the 721 bubbles peak around 6 kpc
(Fig.~\ref{dist}, {\it left} panel). Their distances to the Sun range
from 0.5~kpc to $\sim$ 20~kpc with two peaks at $\sim$ 4 kpc and
$\sim$ 10 kpc (Fig.~\ref{dist}, {\it middle} panel). Less bubbles are
found with distances around $\sim$ 7 kpc from the Sun. Because the
kinematic distance ambiguity problem\footnote{For any objects in the
  inner Galaxy, two possible kinematic distances correspond to a same
  observed line-of-sight velocity, see e.g. ~\citet{rjh+09} for a
  detail.} is hard to solve. We calculated the physical effective
radius for each bubble and found most of them are less than 5 pc
(Fig.~\ref{dist}, {\it right} panel).

\subsection{Survey data of molecular gas}
\label{cogas}

As shown in Sect.~\ref{velocity}, the selected MWP bubbles are
concentrated in the inner Galactic plane. We measured the mean
apparent effective radius of these bubbles to be about
2$\arcmin$. Therefore, in order to explore the gaseous environment of
these bubbles, a systematic survey of molecular line(s) covering the
inner Galaxy with sufficient angular resolution (better than
2$\arcmin$) and sampling is required.

\begin{figure*}
\centering\includegraphics[width=0.9\textwidth]{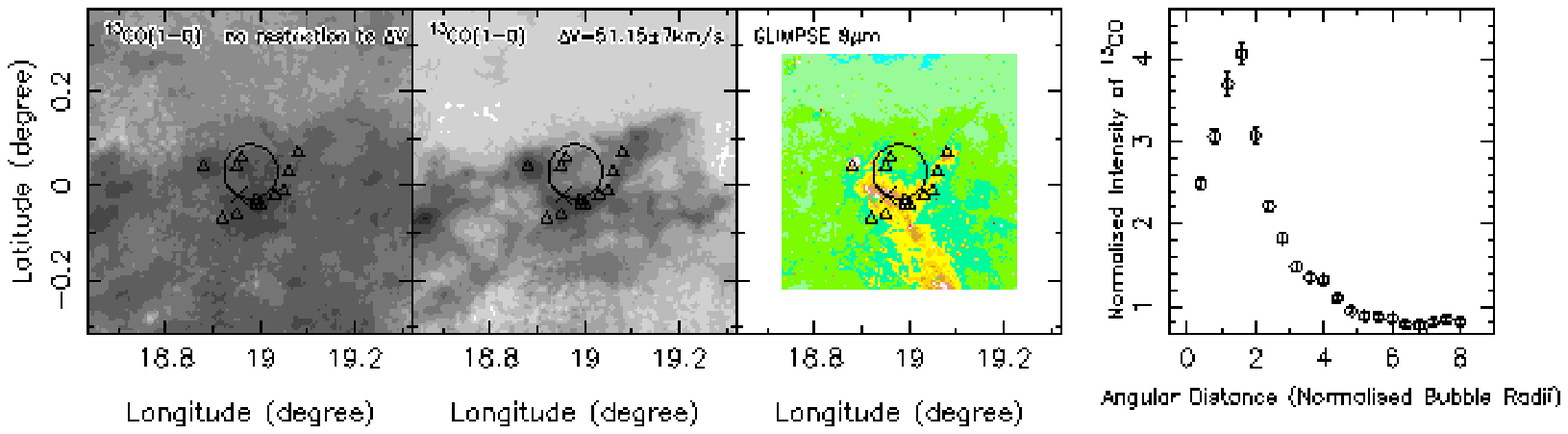}\\
\centering\includegraphics[width=0.9\textwidth]{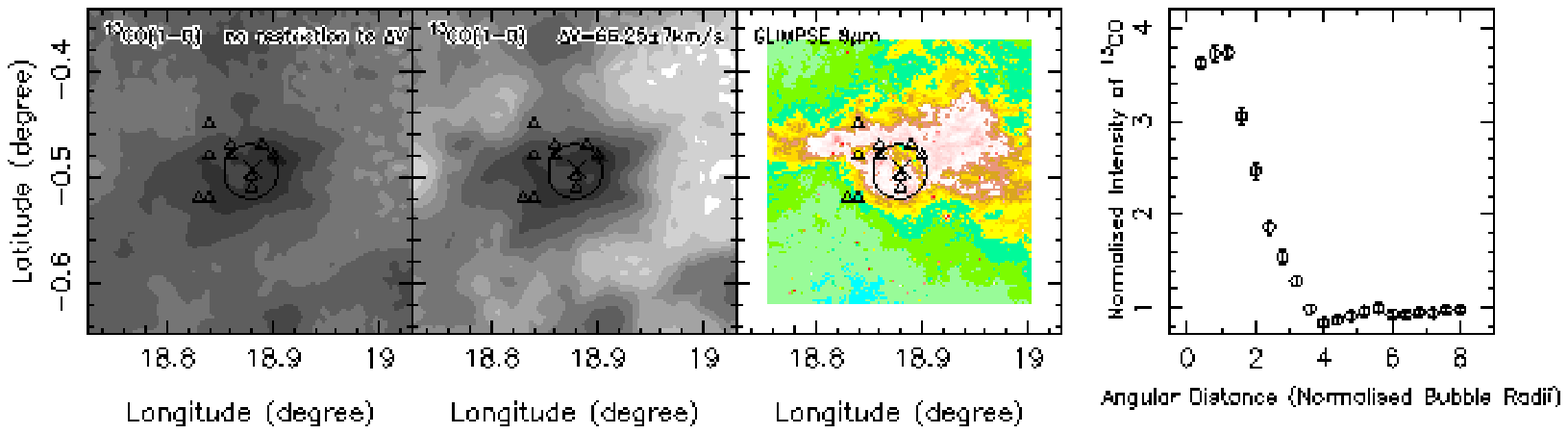}
\caption{{\it Upper left}: velocity integrated maps of $^{13}$CO(1-0)
  for the interstellar bubble MWP1G018980+000304. The integration is
  performed for all the velocity range of $^{13}$CO(1-0) ({\it Left
    first}), and for the bubble-associated velocity range only ({\it
    Left second}). {{\it Left third}}: the GLIMPSE 8~$\mu$m map for
  the same observed regions. {\it Upper right}: normalised intensity
  of $^{13}$CO(1-0) emission as a function of the angular distance to
  the bubble center (see Sect.~\ref{aarp} for a detail). The error
  bars are estimated via the standard error of the mean. In the upper
  left images, the black circles indicate the apparent size of the
  bubble measured by the effective radius~\citep{spk+12}. The black
  crosses show the central position of RRL observations. The black
  triangles indicate the molecular clumps identified
  by~\citet{rjj+09}. {\it Lower panels}: the same as {\it upper
    panels}, but for the interstellar bubble MWP1G018879$-$004949.}
\label{test}
\end{figure*}

The molecular gas in our Milky Way has been widely studied by the
efforts of the Columbia Galactic survey of $^{12}$CO~\citep[][and
  reference therein]{dht01}, the UMASS Stony-Brook (UMSB) survey of
$^{12}$CO~\citep{umsb}, the NANTEN Galactic plane survey of
$^{12}$CO~\citep{mf04}, the FCRAO $^{12}$CO survey of the Outer
Galaxy~\citep{mbs+98}, the Bell Laboratories survey of
$^{13}$CO~\citep{lskm01}, and the Galactic ring survey of
$^{13}$CO~\citep[][hereafter the GRS]{grs06}. The Columbia survey
covers the entire Galactic plane, but the beam size is as coarse as
8.7$\arcmin$. The UMSB survey covers the first Galactic quadrant with
a beam size of 45$^{\prime\prime}$, but the sampling is
3$\arcmin$. The FCRAO survey of the Outer Galaxy and the NANTEN survey
do not overlap with the {\it Spitzer}/GLIMPSE survey. The Bell-Lab
survey covers the first Galactic quadrant, but the angular resolution
(103$^{\prime\prime}$) and the sampling (3$\arcmin$) are not
qualified.

The GRS covering the Galactic longitude range from 18$^\circ$ to
55.7$^\circ$ and the latitude range of $|b| \leqslant$ 1$^\circ$ meets
our requirements by having an angular resolution of
46$^{\prime\prime}$ and a sampling of
22$^{\prime\prime}$. Additionally, the smaller optical depth of
$^{13}$CO(1-0) performs as a much better tracer of the column density
than $^{12}$CO(1-0) and results in narrower line widths, making it
possible to separate the blended lines from distinct clouds with close
velocities \citep{grs06}. Furthermore, the GRS data of $^{13}$CO(1-0)
have been well explored. \citet{rjj+09} identified 829 molecular
clouds (with sizes of 20$-$60~pc) and 6\,124 molecular clumps (with
sizes of 3$-$20~pc) by using its data.

In the sky coverage of the GRS, 362 interstellar bubbles with
associated HII regions remain, and 346 out of the 362 bubbles have
known distances. The 362 bubbles and the $^{13}$CO survey data of the
GRS, constitute the basis of our following statistical study.

\section{ The gaseous environment of {\it Spitzer} interstellar bubbles}

To study the gaseous environment of the {\it Spitzer} interstellar
bubbles, the first step is to extract the gas components which are
{\it truly} associated with the bubbles. However, it is not
straightforward, because the observed spectra of $^{13}$CO(1-0) toward
a bubble are always contaminated by the emission of
foreground/background molecular gas. As discussed in
Sect.~\ref{velocity}, the line-of-sight velocity of a bubble V$_{bub}$
is taken as that of the correlated H\,II region. A similar velocity is
expected for the associated gas components encompassing the bubble. In
this work, we take $\delta$V = 7 km~s$^{-1}$ as the allowance, which
is a commonly adopted value for the velocity uncertainty of massive
star forming regions in the Galaxy. The $^{13}$CO emissions with
velocity between V$_{bub}$-$\delta$V and V$_{bub}$+$\delta$V were
integrated, and taken as the gas components associated with the
bubbles. Different values ($\delta$V = 9 km s$^{-1}$ and $\delta$V = 5
km s$^{-1}$) are also used for the calculation. We found negligible
differences in the conclusions by the following statistical analysis.

To show the advantage by constraining the velocity difference between
the bubbles and the gas components to solve the projection effect, we
compare the results calculated by the integration of the gas
components through the whole velocity range and our defined
``bubble-associated'' range in Fig.~\ref{test}. The result implies the
necessity. The two panels at the top left corner are for the bubble
MWP1G018980+000304, and the two panels at the lower left corner are
for the bubble MWP1G018879$-$004949. It is obvious that, after making
the constraint, the features shown in the velocity integrated maps of
$^{13}$CO(1-0) are better agreed with the arc-shaped morphology of the
bubbles (see the GLIMPSE 8$\mu$m map in the {\it left third} panels.

\subsection{The mean azimuthally averaged radial profile of molecular
  gas}
\label{aarp}

The gas distribution around the {\it Spitzer} interstellar bubbles is
first investigated by the azimuthally averaged radial profile of
molecular gas. The profile is calculated as follows. With respect to
the bubble center, the surrounding regions are separated into many
rings with equal radial size. For each ring, the mean surface
intensity and its standard error are calculated and normalised to the
mean surface intensity of $^{13}$CO emission between the angular
distances of two to eight effective bubble radii. The angular distance
relative to the bubble center is expressed in units of the effective
radius of the bubble. Hence, for the uniformly distributed molecular
gas, the derived normalised intensity of $^{13}$CO emission is equal
to 1, whatever the angular distance is. For the non-uniform
distribution, the higher value of the normalised intensity, the higher
degree of the gas accumulation at the corresponding position(s).

The calculated azimuthally averaged radial profiles for the bubbles
MWP1G018980+000304 and MWP1G018879$-$004949 are shown as examples in
the right two panels of Fig.~\ref{test}. The profile for
MWP1G018980+000304 peaks at $\sim$ 1.5 effective bubble radii, while
that for the bubble MWP1G018879$-$004949 is at $\sim$ 1.0 effective
bubble radius. These results reveal that the over density of molecular
gas is close to the boundaries of the two bubbles.

\begin{figure}
\centering\includegraphics[width=0.45\textwidth]{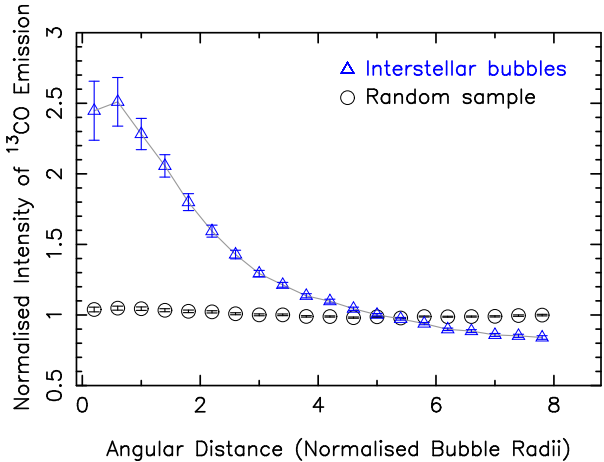}
\caption{Mean azimuthally averaged radial profiles of molecular gas
  for the 309 MWP bubbles (open triangles) and for the random sample
  (open circles). Error bars are estimated via the standard error of
  the mean.}
\label{final}
\end{figure}

The same calculations were then applied to all of the 362 MWP bubbles
which have been selected in Sect.~\ref{cogas}. However, some bubbles
have to be discarded since the velocity coverage of the GRS spectra is
from $-$5 km~s$^{-1}$ to 135 km~s$^{-1}$, out of the selection
criteria (between V$_{bub}-$ $\delta$V and V$_{bub}+$ $\delta$V). Some
other bubbles are located close to the boundary of the GRS sky
coverage, their azimuthally averaged radial profiles cannot be derived
at larger angular distances. Hence, they were also skipped. Finally,
we obtained the azimuthally averaged radial profiles for 309 MWP
bubbles in total. They were then averaged to generate the mean
molecular gas profile of the interstellar bubbles, as presented in
Fig.~\ref{final}. For comparison, the mean profile for a random sample
of 3\,090 bubbles (ten time larger in quality than that of the
selected MWP bubbles) is shown in the same plot. The random sample is
not from any real measurements, but generated artificially. The
parameters (effective radius, coordinates and velocity) of the random
sample are randomly created, but constrained to have similar
distributions of the effective radius, coordinates and velocity to
those of the 309 MWP bubbles.

A clear excess in the normalised intensity of $^{13}$CO at angular
distances less than about three effective bubble radii is visible. A
peak appears at angular distance of about one effective bubble radius,
indicating the presence of gas over density in the immediate periphery
of the bubbles. At larger angular distances, the normalised intensity
of $^{13}$CO gradually falls and reaches the values close to 1. In
comparison, the mean profile always stays around 1 for the random
sample. It implies that the gas component is non-uniformly distributed
around the MWP interstellar bubbles. The intensity is statistically
higher-than-average at smaller effective bubble radii, especially for
the regions close to the bubble rims, and at larger angular distances,
the intensity is much lower. This result for the gas component
resembles the projected distribution of MYSOs around the interstellar
bubbles found by \citet{tumm12} and \citet{ksb+12}.

\begin{figure}
\centering\includegraphics[width=0.45\textwidth]{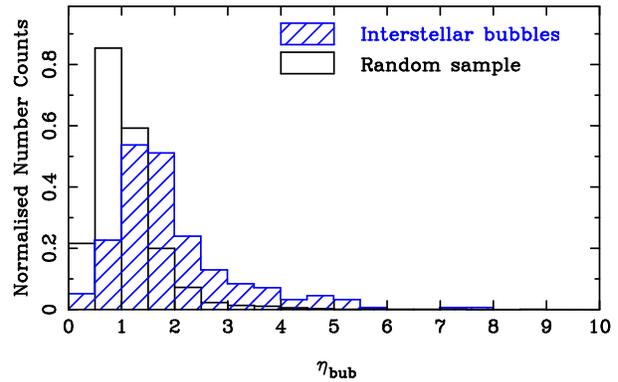}
\caption{Distributions of $\eta_{bub}$ for the 309 MWP bubbles (blue)
  and the random sample (black), where $\eta_{bub}$ is the ratio of
  the normalised intensity of $^{13}$CO integrated between the angular
  distances of 0.5 and 2.0 effective bubble radii to that between 6.0
  and 7.5 effective bubble radii.}
\label{ratio}
\end{figure}
\begin{figure}
\centering\includegraphics[width=0.45\textwidth]{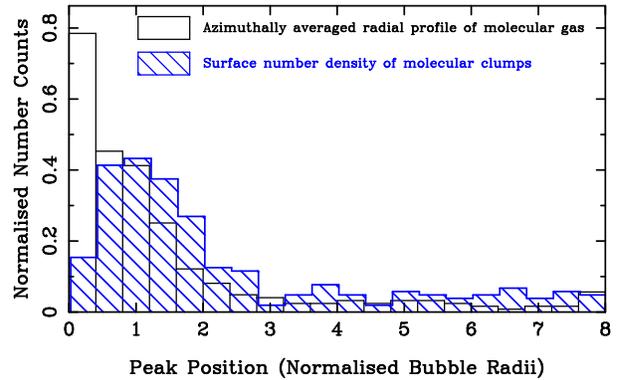}
\caption{Distributions of the peak positions of the azimuthally
  averaged radial profile (black) and the surface number density of
  molecular clumps (blue) for the 309 MWP Bubbles.}
\label{ratio2}
\end{figure}

We note that the averaged profile places emphasis on showing the mean
statistical properties of the gas distribution for the overall
bubbles, but loses sight to individual bubble. Therefore, we calculate
the accumulation parameter $\eta_{bub}$, which is defined as the ratio
of the normalised $^{13}$CO intensity integrated between the angular
distances of 0.5 and 2.0 effective bubble radii to that between 6.0
and 7.5 effective bubble radii. The larger the $\eta_{bub}$, the
higher degree of the gas accumulation near the
bubble. Fig.~\ref{ratio} shows that the gas accumulation degree is
diverse in bubbles. 187 ($\sim$61\%) bubbles are found with
$\eta_{bub}$ $>$ 1.5. We present the distribution of the peak position
in Figure~\ref{ratio2}. Although the majority peak at angular
distances less than two effective bubble radii, some do have peaks at
larger angular distances, indicating the gas over density at the
periphery of the interstellar bubbles does not occur in every case.

\subsection{The surface number density of molecular clumps}

The azimuthally averaged radial profile reveals the statistical over
density of molecular gas near the bubble rims. However, the dense
molecular gas concentrates in molecular clumps, hence the calculation
of the mean surface intensity dilutes these structures and may
underestimate the significance. 6\,124 molecular clumps with $^{13}$CO
emission in the GRS have been identified by \citet{rjj+09}. This
enables us to make a statistical study of the distribution of
molecular clumps around the bubbles.

\begin{figure}
\centering\includegraphics[width=0.45\textwidth]{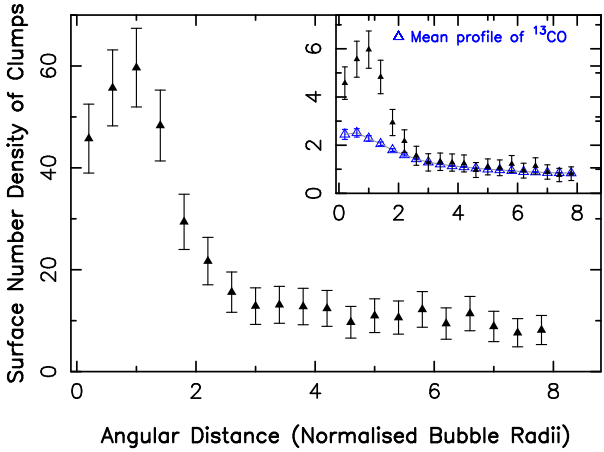}
\caption{Surface number density of molecular clumps as a function of
  the effective bubble radii. Error bars are determined via Poison
  statistics. The mean azimuthally averaged radial profile of
  molecular gas (blue, see Fig.~\ref{final}) is compared in the inset.
}
\label{boson}
\end{figure}

As the method described above, the ``bubble-associated'' clumps were
first selected out by the velocity criteria of $\delta
V=7$~km~s$^{-1}$. Then instead of calculating the mean radial
intensity of $^{13}$CO ring by ring, we estimated the surface number
density of clumps in each ring area (number count of clumps divided by
the ring area\footnote{Here, the ring area is defined as $\pi
  (R_{out}^2-R_{in}^2)$, $R_{out}$ and $R_{in}$ are the outer and
  inner radius of the ring, respectively, in unit of the effective
  bubble radii}). The average distribution for all the 309
investigated bubbles is shown in Fig.~\ref{boson}. Prominent excess
appears at angular distances less than two effective bubble radii. A
significant peak is found near one effective bubble radius at about
7.7 $\sigma$ level. The surface number density of clumps falls sharply
from one effective bubble radius to larger angular distances, and
reaches a background value of about 10 clumps per unit area. In the
inset of Fig.~\ref{boson}, we divide the surface number density by the
background value of 10, and compare with the result calculated by the
method of the azimuthally averaged radial profile shown in
Fig.~\ref{final}. Obviously, the peak revealed by the surface number
density method is more pronounced.

\begin{figure}
\centering\includegraphics[width=0.45\textwidth]{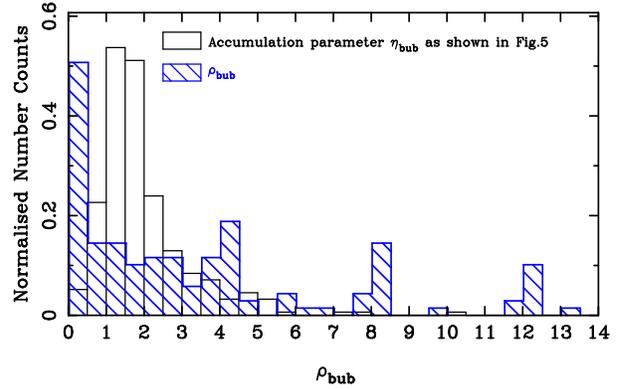}
\caption{Distribution of the accumulation parameter $\rho_{bub}$ for
  the bubbles (blue), where $\rho_{bub}$ is the ratio of the surface
  number density of molecular clumps between the angular distances of
  0.5 and 2.0 effective bubble radii to that between 6.0 and 7.5
  effective bubble radii. The accumulation parameter $\eta_{bub}$ for
  the azimuthally averaged radial profile of molecular gas is compared
  (black, see Fig.~\ref{ratio} and Sect.3.1).}
\label{ratio3}
\end{figure}

Similar with the analysis in Sect.~\ref{aarp}, we define an
accumulation parameter $\rho_{bub}$ as the ratio of the surface number
density of molecular clumps between the angular distances of 0.5 and
2.0 effective bubble radii to that between 6.0 and 7.5 effective
bubble radii. The distribution of $\rho_{bub}$ is shown in
Fig.~\ref{ratio3}. 60\% of the bubbles are found with $\rho_{bub} >$
1.5 and some of the bubbles has accumulation parameter $\rho_{bub} <$
0.5, indicating barely no associated molecular clumps can be
identified in the GRS data. The peak position of the surface number
density of molecular clumps is calculated and compared in
Fig.~\ref{ratio2}.

\subsection{The cross-correlation of MWP bubbles and GRS molecular
  clumps}
\label{cross}

The surface number density of molecular clumps overcomes the dilution
of gas accumulation led by the method of radial intensity profile and
confirms the gas over density around the bubble rims. To learn the
uncertainty and the significance of the over density more
independently, we further check the angular cross-correlation function
of the MWP bubbles and the GRS molecular clumps.

To calculate the angular cross-correlation function, we adopt the
Landy-Szalay estimator~\citep{ll93}:
\begin{equation}
    \omega(\theta)=\frac{N_{DD}(\theta)-2N_{DR}(\theta)+N_{RR}(\theta)}{N_{RR}(\theta)},
\label{eq1}
\end{equation}
where $\theta$ is the angular separation between two objects,
N$_{i,j}(\theta)$ ($i$ = $D$ or $R$, $j$ = $D$ or $R$) is the
normalised number of pairs between sample $i$ and sample $j$ with
angular separation of $\theta$. Subscripts $D$ and $R$ represent the
data sample and the random sample, respectively. For two different
data sets, Eq.~\ref{eq1} can be modified to:
\begin{equation}
    \omega(\theta)=\frac{N_{D_1D_2}(\theta)-N_{D_1R_1}(\theta)-N_{R_1D_2}(\theta)+N_{R_1R_2}(\theta)}{N_{R_1R_2}(\theta)},
\label{eq2}
\end{equation}
as in \citet{bah+11},~\citet{tumm12} and \citet{ksb+12}. In this work,
the subscripts $D_1$ and $D_2$ indicate the MWP bubble and the GRS
molecular clump sample, respectively. The subscripts $R_1$ and $R_2$
indicate the random samples for the bubbles and molecular clumps.
%{ N$_{i,j}(\theta)$} was normalized, by dividing by
%$\Sigma_\theta$N$_{i,j}(\theta)$.

In matching the pairs between the bubbles and the clumps, we only
consider the objects whose velocity difference is less than $\delta
V=7$~km~s$^{-1}$. The random sample of the bubbles is randomly
generated, but with similar distributions of coordinates, velocity and
effective radius to those of the 309 MWP bubbles, and 32 times larger
in quantity. The random sample of the molecular clumps was generated
the same way to those of the 6\,124 GRS molecular clumps. The
bootstrap re-sampling was used to estimate the errors. 100 bootstrap
iterations were made in our analysis.

\begin{figure}
\centering\includegraphics[width=0.45\textwidth]{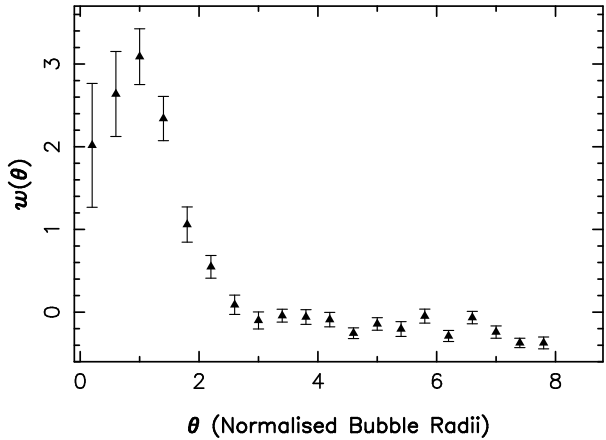}
\caption{Angular cross-correlation function $\omega$($\theta$) for the
  MWP bubbles and the GRS molecular clumps. The error bars were
  estimated by the method of bootstrap re-sampling, and given at the
  1$\sigma$ level.}
\label{tcros}
\end{figure}

We show the cross-correlation result of the MWP bubbles and the GRS
molecular clumps in Fig.~\ref{tcros}. Similar features are found to
that derived by the surface number density of molecular clumps. A
pronounced peak at 9.2$\sigma$ level appears near one effective bubble
radius. There is a prominent excess of the angular cross-correlation
function $\omega$($\theta$) at the angular distances less than about
two effective bubble radii, implying that the probability of finding
the GRS molecular clumps around the bubble rims is significantly
higher than in the outer regions. At angular distances larger than
about three effective bubble radii, $\omega$($\theta$) stays around
0.0, indicating that the molecular clumps at larger distances are not
correlated with bubbles.

\begin{figure}
\centering\includegraphics[width=0.45\textwidth]{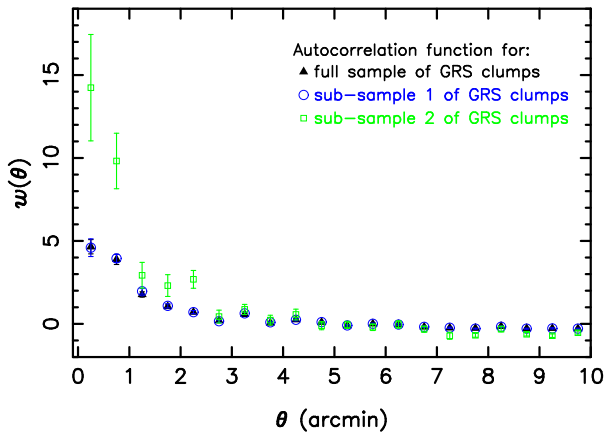}\\
% \centering\includegraphics[width=0.45\textwidth]{tpcc_bubble.ps}
\caption{Results of autocorrelation functions $\omega$($\theta$) for
  the full sample of the GRS molecular clumps (black filled
  triangles), the sample of the bubble-unassociated clumps (sub-sample
  1, blue open circles) and the bubble-associated clumps (sub-sample
  2, green open squares).}
\label{tcros2}
\end{figure}

Possible intrinsic clustering of molecular clumps at a scale close to
the typical angular sizes of the bubbles could contaminate the
estimates of the over density of molecular clumps near the bubble
rims. We inspect this effect as \citet{tumm12} and \citet{ksb+12}. The
6\,124 GRS molecular clumps are separated into two
sub-samples. Sub-sample 1 contains 3\,068 clumps lying outside three
effective bubble radii of any of the 5\,106 MWP bubbles and that are
essentially unassociated with any bubbles. Sub-sample 2 contains the
bubble-associated clumps, which includes 492 molecular clumps lying
within two effective bubble radii from one of the 309 MWP bubbles, and
having velocity difference less than $\delta V=7$~km~s$^{-1}$. The
auto-correlation functions (see Eq.~\ref{eq1}) for the full sample of
the GRS molecular clumps, sub-sample 1 and sub-sample 2 were
calculated and presented in Fig.~\ref{tcros2}.

\begin{table*}
  \caption{Properties for the clump/MYSO-associated and unassociated
    MWP bubbles. The T-statistic significance for the properties of
    the clump-associated/-unassociated and
    MYSO-associated/-unassociated bubbles are given in Column (3) and
    (6), respectively. A small value of the significance indicates
    that the two investigated samples have significantly different
    means.}
\begin{center}
\begin{tabular}{rlcccccc}
  \hline
  \hline
  &                 & clump-associated  &  clump-unassociated  & T-statistic   & MYSO-associated  &  MYSO-unassociated  &  T-statistic\\
  &                 &     bubbles       &   bubbles            & significance  & bubbles          &   bubbles  & significance \\
  &                 &       (1)         &     (2)              & (3)            &   (4)          &   (5)      &   (6)        \\
  \hline
  \multicolumn{2}{l}{ mean effective radius} &     &  &   &  &    &    \\
  &{\it apparent} (arcmin)    & 1.49 $\pm$ 0.08    & 0.89 $\pm$ 0.06& { $<$ 0.1\%}   & 1.41 $\pm$ 0.12    & 1.15 $\pm$ 0.05  &  { 1.8\%} \\
  &{\it physical} (pc)        & 2.90 $\pm$ 0.18    & 2.26 $\pm$ 0.15& {  2.0\% }      & 3.17 $\pm$ 0.30    & 2.36 $\pm$ 0.11 &  { 0.2\%} \\
  \hline
  \multicolumn{2}{l}{mean effective thickness} & &    & &  &       &        \\
  &{\it apparent} (arcmin)   & 1.44 $\pm$ 0.06    & 0.95 $\pm$ 0.06 & { $<$ 0.1\%}   & 1.37 $\pm$ 0.10    & 1.18 $\pm$ 0.05  &  { 4.6\%} \\
  &{\it physical} (pc)       & 2.80 $\pm$ 0.16    & 2.40 $\pm$ 0.15 &  { 13.5\% }     & 3.08 $\pm$ 0.28    & 2.39 $\pm$ 0.10 &  { 0.5\%} \\
  \hline
  \multicolumn{2}{l}{ ionizing photo rate} & &   &   &  &        &     \\
  & log($N_{Ly}$)   & 47.54 $\pm$ 0.09    & 47.01 $\pm$ 0.13        &   { $<$ 0.1\%}    &  { 47.66 $\pm$ 0.11}    & { 47.13 $\pm$ 0.09}  & { $<$ 0.1\%} \\
  \hline\hline
\end{tabular}
\end{center}
\label{para}
\end{table*}

The overall looks of the auto-correlations for the full sample and the
sub-sample 1 resemble each other. Small excesses can only be found at
angular distance less than $\sim$ 1.0$\arcmin$. This indicates the
intrinsic clustering of the molecular clumps for the full sample and
sub-sample 1 only occur at small scales, less than 1.25$\arcmin$, the
mean angular radius for the 309 MWP bubbles. The auto-correlation for
the sub-sample 2 (bubble-associated clumps) behaves differently,
showing a more significant peak at the angular separation of about
0.3$\arcmin$, indicating a strong clustering on that scale. A
secondary peak which is much weaker appears on the scale of $\sim$
2.3$\arcmin$. This is possibly the signature of the molecular clumps
located on either side of bubbles. In all, through the comparisons
between the auto-correlation of different samples and the results
shown in Fig.~\ref{tcros} and 10, it seems that the intrinsic
clustering of the molecular clumps may have, but very weak effects to
the significance of the gas over density around the interstellar
bubble rims.
% shown in the correlation function.

\section{Discussions}

The analysis presented above with three different methods give
consistent results, showing apparent over density of molecular gas
near the interstellar bubble rims. In the following, we investigate
the properties of the clump-associated and clump-unassociated bubbles,
and their differences. By combining the results, we try to determine
whether there are any evidence of triggered material accumulation and
star formation by the expansion of bubbles.

\subsection{The properties of clump-associated and clump-unassociated
  bubbles}

\begin{figure}
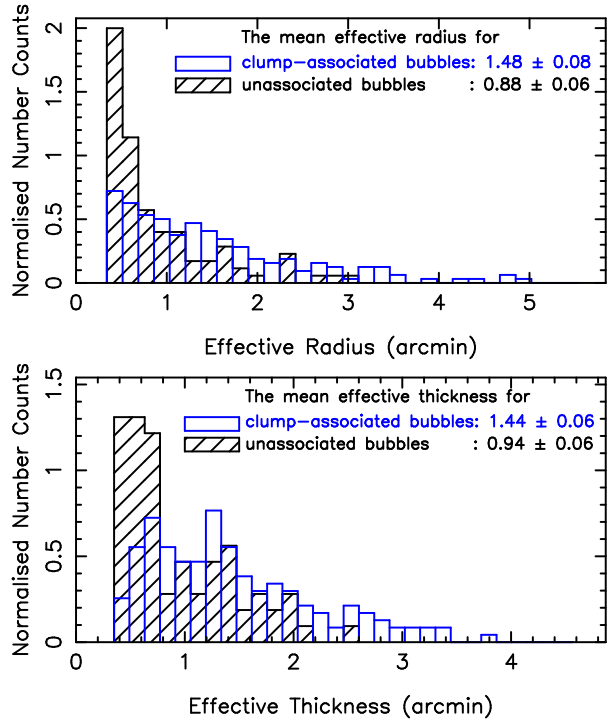

  \centering\includegraphics[width=0.45\textwidth]{b_effr.ps}
  \centering\includegraphics[width=0.45\textwidth]{b_thick.ps}
  \caption{Distributions of the effective radius ({\it upper panel})
    and the effective thickness ({\it lower panel}) for the MWP
    bubbles with and without associated molecular clumps.}
\label{diff}
\end{figure}

As shown in Fig.~\ref{tcros}, the probability of finding GRS molecular
clumps within two effective bubble radii is significantly higher than
in the outer regions. Hence, we define a ``clump-associated bubble''
as a bubble that has associated molecular clump(s) within two
effective bubble radii and the velocity difference between the
bubble and the clump(s) is less than $\delta V=7$~km~s$^{-1}$. 184
($\sim$ 60\%) out of the 309 MWP bubbles meet the criteria and 117
bubbles ($\sim$ 38\%) have more than one matched clump. We define the
bubbles which have no associated clumps within three effective bubble
radii as ``clump-unassociated''. This group contains 101 ($\sim$ 33\%)
MWP bubbles. The clumps located between two and three effective bubble
radii are less significantly correlated with the bubbles and these
cases cannot be clearly classified and are not included in the
analysis. In the following, we discuss the properties of the physical
sizes and the central ionizing sources of the clump-associated and
clump-unassociated bubbles.

The distributions of the effective radius and thickness\footnote{ The
  effective thickness is defined as
  $(R_{out}r_{out})^{0.5}-(R_{in}r_{in})^{0.5}$, where $R_{in}$ and
  $R_{out}$ are the inner and outer semi-major axes of a bubble, and
  $r_{in}$ and $r_{out}$ are the inner and outer semi-minor axes of a
  bubble, respectively (Simpson et al. 2012).} of the clump-associated
and unassociated bubbles are compared (see Fig.~\ref{diff}) and listed
in Table~\ref{para}. Clearly, the clump-associated bubbles tend to be
larger and thicker in physical sizes, and the differences are always
significant, e.g. the probability that the clump-associated and
clump-unassociated bubbles are drawn from populations with the same
mean physical effective radius is only 2.0\%. This contradicts with
the result revealed by the MYSOs around 322 { CH06\&07}
bubbles~\citep{tumm12}. \citet{tumm12} found the MYSO-associated
bubbles tend to be smaller and thinner than the MYSO-unassociated
bubbles. We attribute this discrepancy to the incompleteness of the
CH06\&07 catalogue. \citet{cpa+06} suggested that the incompleteness
of the CH06\&07 bubble catalogue is $\sim$ 50\%. The recently released
bubble catalogue~\citep{spk+12} that we used in this work has ten
times more bubbles than in the CH06\&07 catalogue.

For a further inspection of this contradiction, we cross match the 309
MWP bubbles with the MYSO sample of \citet{uhl+08}. According to
\citet{ksb+12}, if the angular separation between a MYSO and a bubble
is less than 1.6 effective bubble radii, they are taken as
coincidence. 102 ($\sim$ 33\%) of the 309 MWP bubbles are then found
to be MYSO-associated. The bubbles that have no MYSOs within two
effective bubble radii are taken as
MYSO-unassociated\citep{ksb+12}. 195 ($\sim$ 63\%) of the 309 MWP
bubbles are selected. The mean values for the effective radius and
thickness are then calculated and compared in Table~\ref{para}. With
the more complete bubble sample, and the same MYSO sample with
\citet{tumm12}, we claim that both of the MYSO-associated and the
clump-associated bubbles have larger effective radius and thickness
than the un-associated ones.

\citet{dwwp09} suggested that the bubbles with smaller radii and
thinner shells should be younger than the larger and thicker
ones. Hence, the mean values of the effective bubble radii and
thickness could be rough indicators of the bubble ages. From
Table~\ref{para}, we do see such differences between the
clump-/MYSO-associated and the clump-/MYSO-unassociated bubbles in
their mean physical effective radii and thicknesses. Therefore we
speculate that the age of bubbles might be a factor for a bubble to
have associated molecular clumps or MYSOs.

In addition, we also try to search for the differences of their
central ionizing source(s). We calculated the ionizing photon
rate~\citep{mesz76}:
\begin{equation}
   N_{Ly}~=~7.5~\times~10^{46}~F_\nu~\nu^{0.1}~d^2~T_e^{-0.45}~[photos~s^{-1}]
\label{eq4}
\end{equation}
where $F_\nu$ is the flux density in Jy, derived from the MAGPIS
survey~\citep{hbw+06}, $\nu$ is the observed frequency in GHz, $d$ is
the bubble distance to the Sun in kpc, which should be consistent with
the distance of central ionizing source(s), and $T_e$ is the electron
temperature in 10$^4$ K. Here, we adopted the typical electron
temperature 8500 K for the H\,II regions.
% The distribution of the ionized photo rate is shown in
% Fig.~\ref{lumi}.
Different values are found for the mean ionizing photon rate
log($N_{Ly}$): 47.54 $\pm$~0.09 for the clump-associate bubbles and
47.01 $\pm$ 0.13 for the bubbles without clumps. The probability that
the two results are taken from the populations with the same mean
ionizing photon rate is less than 0.1\%. Hence, the luminosity of
central ionizing source(s) seems to be another factor for a bubble to
have associated molecular clumps.

The reaction of molecular gas to the expansion of H\,II regions is
complex and determined by several factors: the luminosity of central
ionizing source(s), the ability of central ionizing source(s) to
ionize the surrounding gas, which depends on the ambient gas density
and the strength of accretion, and the ability of H\,II region to
drive out the ambient neutral gas, which depends on the escape
velocity of the whole system with respect to the sound speed of the
ionized gas~\citep[see][for detail]{deb12}. As shown above, both the
age of the interstellar bubbles and the luminosity of the central
ionizing source(s) seem to have certain impacts on a bubble to have
associated molecular clump(s)/ MYSO(s). Additionally, the properties
of the ambient gaseous environment, e.g. the density field and the
velocity field, which might also play an important role, deserves more
attention in the future high-resolution observations.

\subsection{The properties of bubble-associated and
  bubble-unassociated molecular clumps}
\label{clumps}

In this part, we simply switch our targets from the MWP interstellar
bubbles to the GRS molecular clumps. We compare the properties of
``bubble-associated'' and ``bubble-unassociated'' clumps. With the
same velecity selection criteria. 492 out of the 5\,106 GRS clumps are
found to be associated with the 309 interstellar bubbles. For all of
the clumps which are located outside the three effective radii from
any of the MWP bubbles, we define them as the bubble-unassociated
clumps.

\begin{figure}
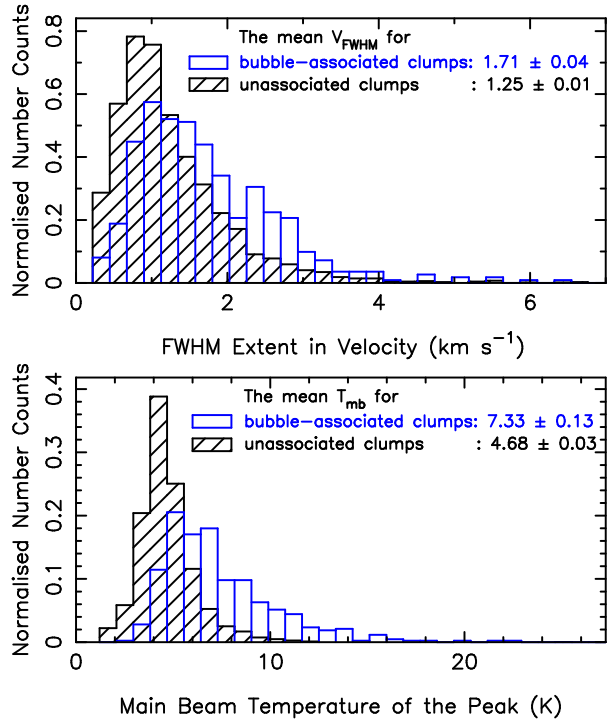

\centering\includegraphics[width=0.45\textwidth]{dv_mc.ps}
\centering\includegraphics[width=0.45\textwidth]{Tmb_mc.ps}
\caption{Distributions of the FWHM extent in velocity ({\it upper
    panel}) and the main beam temperature of the peak ({\it lower
    panel}) for the bubble-associated and unassociated molecular
  clumps.}
\label{diff2}
\end{figure}

We find that the mean velocity FWHM extent \citep[V$_{FWHM}$, defined
as 2.35 times the velocity dispersion of the molecular clump, given
by][]{rjj+09} is 1.71 $\pm$~0.04~km~s$^{-1}$ for the bubble-associated
clumps, and 1.25~$\pm$~0.01~km~s$^{-1}$ for the bubble-unassociated
clumps. The peak main beam temperature (T$_{MB~peak}$) for the two
groups are 7.33~$\pm$~0.13~K and 4.68~$\pm$~0.03~K, respectively~(see
Fig.~\ref{diff2}). It is clear that the bubble-associated molecular
clumps tend to have larger T$_{MB~peak}$ and V$_{FWHM}$, and have
larger peak intensity and larger peak H$_2$ column density
consequently, namely, they are brighter and denser. \citet{dcb07}
examined the impact of the external ionizing irradiation on a
turbulent molecular cloud numerically. They found that the molecular
cloud exposed to the ionizing radiation field tends to be denser, and
forms about twice as many young stars in comparison to the control
model. A natural explanation is that the shock driven by the ionizing
feedback will sweep up and compress the surrounding neutral gas,
hence, the molecular clumps influenced by the feedback tend to be
brighter and denser. The differences between the bubble-associated and
bubble-unassociated molecular clumps might reflect the influence of
ionizing feedback from central massive star(s)/star cluster that
creates the bubbles.

\subsection{Triggered material accumulation and star formation by the
  expansion of bubbles?}

The observational over density of molecular gas near the bubble rims
is consistent with the scenario that the bubble expansion has a
prominent impact on the ambient gas and triggers the material
accumulation~\citep[see][]{dsa+10}. But it can also be interpreted as
a fact that the bubbles tend to form in the high density regions of
molecular gas.

\setlength{\tabcolsep}{1.2mm}
\begin{table}
  \caption{Properties for the MYSO-associated and unassociated
    bubble-clump complexes.}
\begin{center}
\begin{tabular}{rlccc}
  \hline
  \hline
  &                 & MYSO-  &  MYSO-   &  { T-statistic }  \\
  &                 &     associated &   unassociated  & { significance}       \\
  % &                 &      complexes   &    complexes                  \\

  \hline
  \multicolumn{4}{l}{mean effective radius}           \\
  &{\it apparent} (arcmin)    & 1.65 $\pm$ 0.15    & 1.33 $\pm$ 0.08 & { 4.3\%}  \\
  &{\it physical} (pc)        & 3.45 $\pm$ 0.40    & 2.53 $\pm$ 0.16 & { 1.8\%}  \\
  \hline
  \multicolumn{4}{l}{mean effective thickness}                  \\
  &{\it apparent} (arcmin)   & 1.54 $\pm$ 0.12    & 1.33 $\pm$ 0.06  &  { 8.6\%}  \\
  &{\it physical} (pc)       & 3.23 $\pm$ 0.35    & 2.52 $\pm$ 0.14  &  { 3.7\%}\\
  \hline
  \multicolumn{4}{l}{ionizing photo rate}                 \\
  & log($N_{Ly}$)   & 47.64 $\pm$ 0.14    & 47.44 $\pm$ 0.12  & { 26.7\%}  \\
  \hline
  \multicolumn{4}{l}{FWHM extent in velocity}                   \\
  & (km~s$^{-1}$)   & 1.81 $\pm$ 0.06    & 1.66 $\pm$ 0.06  & { 7.8\%}  \\
  \hline
  \multicolumn{4}{l}{main beam temperature of the peak}                   \\
  & (K)   & 7.58 $\pm$ 0.18    & 7.27 $\pm$ 0.17   &  { 22.0\%} \\
  \hline\hline
\end{tabular}
\end{center}
\label{pa2}
\end{table}

If the later is correct, one question would be that why the
distribution of the molecular gas peculiarly peaks near the bubble
rims. The bubble rims are traced by the 8~$\mu$m emission, which is
dominanted by the emission from the PAH or clusters of
molecules. These species are excited by the absorption of far-UV
photons leaking from the central ionizing sources and trace the
expanding ionization fronts of H\,II regions. Hence, the bubble rim is
time-dependent, increasing with time. If the bubbles formed in high
density regions of neutral gas, for the spherical or
ring-like~\citep{bw10} morphology of the bubbles, the column density
of molecular gas in the outer regions of the bubbles is expected to be
close to that near the bubble rims, because of the similar integrated
path length penetrating the line of sight. However, this is not the
case as we found in Fig.~\ref{boson}, that the material density falls
rapidly from the angular distance from one effective bubble radius to
larger distances.

Alternatively, the influence of bubble expansion on the ambient gas
can naturally explain the observed properties of gas distribution. The
expansion of bubbles is caused by the higher pressure in the ionized
gas than its surrounding neutral material. This process is accompanied
by the accumulation of neutral material between the ionizing front and
the shock front~\citep[e.g. see][]{dsa+10}. Our results suggest that
some of the bubble-associated clumps are probably triggered by the
collecting process during the bubble expansion.

More intriguing questions are whether the bubble-associated molecular
clumps are able to generate new stars, and whether the bubble
expansion influences the formation of the new stars. In our sample, as
we mentioned above, 184 bubbles have associated molecular
clump(s). They are now named as ``bubble-clump complexes''. We check
their associations with MYSOs.

If the bubble expansion does not influence the star formation in the
ambient dense gas, one would expect no differences between the
properties of the MYSO-associated bubble-clump complexes and the
MYSO-unassociated complexes. However, we found differences in the mean
effective radius and the thickness for the bubbles in different groups
as given in Table~\ref{pa2}. Those differences are around 3$\sigma$
for the physical sizes, and slightly above 2$\sigma$ for the apparent
sizes. The bubbles in the MYSO-associated complex are larger and
thicker, hence, tend to be older. We further evaluate other factors
which may take effect in this process for the two different groups,
such as the mean ionizing photon rate log$N_{Ly}$ of the central
ionizing source(s), the peak main beam temperature and the FWHM extent
in velocity for the clumps, but no significant differences were
found. Therefore we speculate that, as the bubble in the complex
evolves, the relatively younger MYSO-unassociated complexes could
breed MYSOs, and may finally turn to be MYSO-associated. If so, the
MYSO-unassociated complexes are the potential sites of triggering SF
by the feedback of H\,II regions. To verify this hypothesis, detailed
studies on the physical conditions and dynamical properties of the
neutral gas are necessary. High-resolution observations of different
molecular species, e.g. $^{13}$CO, NH$_3$, CH$_3$OH, are required.

\subsection{The fraction of molecular clumps associated with MWP
  bubbles.}

As discussed in section~\ref{clumps}, a bubble-associated clump is
defined as a clump that has one bubble within two effective bubble
radii. To solve the projection effect, their velocity difference
should be less than $\delta V=7$~km~s$^{-1}$. 492 (about 8\%) of the
6\,124 molecular clumps are selected finally. This percentage is the
lower limit, because only 309 of the 1\,637 MWP bubbles in the sky
coverage of the GRS have their line-of-sight velocities measured. If
no constraint was made on the velocity difference, namely, the
projection effect was not considered, this number increases to 788
(about 13\%), that match the 309 interstellar bubbles. This leads to a
significant percentage (about 38\%) of false identification.

To estimate the fraction of molecular clumps associated with MWP
bubbles, we used all the 1\,637 MWP bubbles falling into the sky
coverage of the GRS and made no constraints on the velocity
difference. 1\,980 (about 32\%) of the 6\,124 molecular clumps are
then selected out to have matched MWP bubbles. By assuming that 38\%
of them are mistaken due to the projection effect, the number of
associations is reduced to 1\,236 (about 20\%). This fraction varies
from 16\% to 22\% corresponding to $\delta V=5$~$-$~9~km~s$^{-1}$.

The result seems to be consistent with the previous estimates of the
fraction of MYSOs associated with the interstellar bubbles. By
analyzing the associations of RMS MYSOs and the CH06\&07 bubbles,
\citet{tumm12} estimated the fraction of MYSOs triggered by the
expansion of bubbles is between 14\% and 30\%. By using a more
complete catalogue of bubbles, \citet{ksb+12} obtained a fraction of
$\sim$ 22\% $\pm$ 2\%. With a relative small sample of bubbles, this
fraction was estimated to be 18\% and 20\% by~\citet{dsa+10}
and~\citet{whm10}, respectively. Note however, that the projection
effect was not considered in any of the estimates mentioned
above. Their results might be somehow overestimated.

\section{Conclusions}

We made a statistical study of the gaseous environment of the
interstellar bubbles identified by the Milky Way
Project~\citep{spk+12}. Our conclusions are summarized below:

$\bullet$ By making cross-identification between the MWP {\it Spitzer}
interstellar bubbles and the Galactic H\,II region catalogue compiled
by us, we obtain the velocity information for 818 MWP bubbles. As a
result, 721 bubbles have determined distances, either photometric,
trigonometric or kinematic, which is nearly four times larger than
previous known.

$\bullet$ We used three methods to study the gaseous environment of
MWP bubbles:

(1) The mean azimuthally averaged radial profile of $^{13}$CO
emission, with which we found clear emission excesses at the angular
distances less than three effective bubble radii. A significant peak
appears close to one effective bubble radius, corresponding to the
bubble rims. At larger angular distances, the mean profile gradually
falls and finally reaches the values close to the background level.

(2) The surface number density of molecular clumps is also found to be
enhanced toward the bubble rims. The excess at smaller angular
distances is more prominent than that shown by the mean azimuthally
averaged radial profile of $^{13}$CO emission. A significant peak at a
7.7 $\sigma$ level is found near one effective bubble radius. The
surface number density falls more sharply at larger angular distances.

(3) The cross-correlation function of the MWP bubbles and the GRS
molecular clumps gives consistent results with that of the surface
number density of molecular clumps. The affection of intrinsic
clustering of the GRS clumps is estimated and found to be
negligible. At larger angular distances, e.g. $>$ three effective
bubble radii, the molecular clumps are essentially not correlated with
the interstellar bubbles.

Our results for the distribution of gas component around the bubbles
are consistent with the analysis based on the associations between the
MYSOs and the bubbles. The significant peak near the bubble rims and
the variations of molecular gas along the angular distance make us to
speculate that some of the bubble-associated clumps might be triggered
by the expansion of bubbles.

$\bullet$ About 60\% of the investigated interstellar bubbles have
associated molecular clumps. By comparing the effective radius,
thickness and the ionizing source(s) of the clump-associated and
clump-unassociated bubbles, we speculate that the age and the
luminosity of central ionizing source(s) have certain impacts for a
bubble to have associated molecular clumps.

$\bullet$ For the bubble-clump complexes, we inspected their
associations with MYSOs. The bubbles in the MYSO-associated complex
are found to be larger and thicker, hence relative older. This implies
an evolutionary sequence that the relative younger MYSO-unassociated
bubble-clump complexes might breed MYSOs, and evolve to the phase of
MYSO-associated.

$\bullet$ The fraction of molecular clumps associated with the MWP
bubbles is estimated to be about 20\% after considering the projection
effect.

\section*{Acknowledgements}

We thank the referee Dr. Nicolas Flagey for constructive comments.
LGH would like to thank Dr. L{\'e}pine and Dr. Bronfman for kindly
providing us their data. The authors are supported by the National
Natural Science Foundation (NNSF) of China (10773016 and 11303035),
the Youth Foundation of Hebei Province (A2011205067). XYG is
additionally supported by the Young Researcher Grant of National
Astronomical Observatories, Chinese Academy of Sciences.

 This publication makes use of molecular line data from the Boston
 University-FCRAO Galactic Ring Survey (GRS). The GRS is a joint
 project of Boston University and Five College Radio Astronomy
 Observatory, funded by the National Science Foundation under grants
 AST-9800334, AST-0098562, AST-0100793, AST-0228993, \&
 AST-0507657.This paper made use of information from the Red MSX
 Source survey database at www.ast.leeds.ac.uk/RMS which was
 constructed with support from the Science and Technology Facilities
 Council of the UK.

\bibliographystyle{aa}
\bibliography{bubble}

\begin{appendix}
\section{{\it Spitzer} interstellar bubbles with velocity information}

We cross-identified the MWP bubbles and the Galactic H\,II regions,
818 matched pairs are found and listed in Table~\ref{tab_1}, which is
available in its entirety in the online material, and a portion is
presented below for guidance regarding its form and content.

\scriptsize
\setlength{\tabcolsep}{1.4mm}
\begin{longtable}{lccrrrlrrrrrrlr}
  \caption{\label{tab_1}Matched pairs of MWP bubbles and Galactic
    H\,II regions. Columns 1 to 4 list the MWP name, Galactic
    longitude, Galactic latitude, and effective radius for each
    bubble; Cols 5 and 6 give the Galactic longitude and Galactic
    latitude of the associated HII regions, taken from the reference
    listed in Col. 7; Col. 8 is the line-of-sight velocity for each
    HII region; Cols. 9 to 10 list the trigonometric or photometric
    distance and the uncertainty when available from the reference
    shown in Col. 11; Col. 12 gives the kinematic distance estimated
    with a flat rotation curve with R$_0$~=~8.3~kpc and
    $\theta_0$~=~239~km~s$^{-1}$; Col. 13 is the uncertainty of the
    kinematic distance estimated by considering a velocity uncertainty
    of $\pm$ 7 km s$^{-1}$; Col. 14 gives the solution of the
    kinematic distance ambiguity: kfar - the farther kinematic
    distance is adopted; ktan - the source is located at the
    tangential point; knear - the nearer kinematic distance is
    adopted; 3kpcn/3kpcf - the source is in the near or far 3-kpc arm
    \citep[see e.g.,][]{dt11}; and Col. 15 is the reference.\\
    References: abbr11: \citet{abbr11}; abbr12: \citet{abbr12}; ab09:
    \citet{ab09}; ahck02: \citet{ahck02}; bwl09: \citet{bwl09};
    brba11: \citet{brba11}; brm+09: \citet{brm+09}; bab12:
    \citet{bab12}; bfs82: \citet{bfs82}; bdc01: \citet{bdc01}; bnt89:
    \citet{bnt89}; brof96: \citet{brof96}; ch87: \citet{ch87}; cwc90:
    \citet{cwc90}; dwbw80: \citet{dwbw80}; frwc03: \citet{frwc03};
    gcbt88: \citet{gcbt88}; gg76: \citet{gg76}; gm11: \citet{gm11};
    hze+11: \citet{hze+11}; jd12: \citet{jd12}; kb94: \citet{kb94};
    kjb+03: \citet{kjb+03}; lra+11: \citet{lra+11}; lock79:
    \citet{lock79}; lock89: \citet{lock89}; lock96: \citet{lock96};
    msh+11: \citet{msh+11}; mdf+11: \citet{mdf+11}; noh+11:
    \citet{noh+11}; pdd04: \citet{pdd04}; pco10: \citet{pco10};
    rjh+09: \citet{rjh+09}; rus03: \citet{rus03}; rza+12:
    \citet{rza+12}; srm+09: \citet{srm+09}; sh10: \citet{sh10}; st79:
    \citet{st79}; srby87: \citet{srby87}; srbm10: \citet{srbm10};
    swa+04: \citet{swa+04}; uhl12: \citet{uhl12}; was+03:
    \citet{was+03}; wam82: \citet{wam82}; wwb83: \citet{wwb83};
    wmgm70: \citet{wmgm70}; xmr+11: \citet{xmr+11}.
  } \\
  \hline \hline
  MWP name & l$_{\rm bubble}$ & b$_{\rm bubble}$ & Reff & l$_{\rm H\,II}$ & b$_{\rm H\,II}$ & Ref. &v$_{\rm H\,II}$ & d$_{\rm o}$ & d$_{\rm erro}$ & Ref. & d$_{\rm k}$  & d$_{\rm k-erro}$   & Mark & Ref. \\
  &  $(^\circ)$     & $(^\circ)$     & ($^\prime$) & $(^\circ)$ &  $(^\circ)$  &      &   (km/s)  & (kpc)    &  (kpc)         &  & (kpc) & (kpc)    &      &               \\
  (1)     &   (2)       &  (3)       & (4)    & (5)   & (6)      & (7)   & (8)    & (9)    & (10)         & (11) & (12) & (13)      & (14) & (15)   \\
  \hline
\endfirsthead
\caption{continued.}\\
\hline\hline
  MWP name & l$_{\rm bubble}$ & b$_{\rm bubble}$ & Reff & l$_{\rm H\,II}$ & b$_{\rm H\,II}$ & Ref. &v$_{\rm H\,II}$ & d$_{\rm o}$ & d$_{\rm erro}$ & Ref. & d$_{\rm k}$  & d$_{\rm k-erro}$   & Mark & Ref. \\
  &  $(^\circ)$     & $(^\circ)$     & ($^\prime$) & $(^\circ)$ &  $(^\circ)$  &      &   (km/s)  & (kpc)    &  (kpc)         &  & (kpc) & (kpc)    &      &               \\
  (1)     &   (2)       &  (3)       & (4)    & (5)   & (6)      & (7)   & (8)    & (9)    & (10)         & (11) & (12) & (13)      & (14) & (15)   \\
\hline
\endhead
\hline
\endfoot
MWP1G000127+000485   &  0.127   & +0.048  &  3.58  &    0.09   &  0.01   &   wwb83  &   $-$29.70  &        &        &         &              &            &         &         \\              
MWP1G000140$-$001173   &  0.140   & $-$0.117  &  3.62  &    0.094  & $-$0.154  &  brof96  &    16.00  &        &        &         &              &            &         &         \\              
MWP1G000280$-$004800S  &  0.28    & $-$0.48   &  0.59  &    0.284  & $-$0.478  &  dwbw80  &    19.25  &        &        &         &         7.86 &       0.16 &   knear &   rus03 \\              
MWP1G000515$-$007049   &  0.515   & $-$0.705  &  3.24  &    0.489  & $-$0.668  &  dwbw80  &    17.50  &        &        &         &         7.46 &       0.33 &   knear &   rus03 \\              
MWP1G000530+001800S  &  0.53    & +0.18   &  0.50  &    0.527  &  0.182  &  brof96  &    $-$3.05  &        &        &         &              &            &         &         \\              
\hline\hline
\end{longtable}
\normalsize

\end{appendix}
\label{lastpage}
\end{document}